# Closed-Form Mathematical Representations of Interval Type-2 Fuzzy Logic Systems

**Sherif M. Abuelenin[1], Rabab F. Abdel-Kader[2]**

[1]*Electrical Engineering Department, Port-Said University,*
*Port-Fouad, Port-Said 42526 Egypt.*
*E-mail: s.abuelenin@eng.psu.edu.eg*

[2] *Electrical Engineering Department, Port-Said University,*
*Port-Fouad, Port-Said 42526 Egypt.*
*E-mail: rababfakader@eng.psu.edu.eg*

**Abstract**

Interval type-2 fuzzy logic systems (IT2 FLSs) have a wide range of applications due to their abilities to handle uncertainties compared to their type-1 counterparts. This paper discusses the representation of IT2 FLSs in closed mathematical form. Two novel inference mechanisms are introduced, each of them represents a whole IT2 FLS. The two forms are based on approximating Coupland and John's geometric method and the Nie-Tan method. Having closed-form representations is preferred in the design of control systems, especially when stability analysis is studied. Additionally, the simplicity of the proposed mechanisms, offers an easy way for implementation. Simulation results show that the proposed method perform very closely to other methods. Simulation results are provided.

## 1. Introduction

Type-2 fuzzy sets (T2 FSs) were first proposed by Zadeh in 1975 [1] as an extension to type-1 fuzzy sets (T1 FSs) to better handle uncertainties. The main difference between a T2 FS and a T1 FS is that the membership grades are not crisp numbers, but fuzzy sets (secondary membership functions) [2, 3, 34], making the T2 FS a three-dimensional set. A T2 FS has a footprint of uncertainty (FOU) that represents uncertainties in defining its shape and position. When the secondary membership function (MF) is a unit interval for all the points in the primary membership this set is called an interval type-2 fuzzy set (IT2 FS) [4, 40]. An IT2 FS is totally defined by its FOU because no information is contained in the third dimension [4, 5].

A fuzzy system that utilizes at least one IT2 FS in the antecedent or consequent part of its rule-base is known as an interval type-2 fuzzy logic system (IT2 FLS) [3, 6]. IT2 FLSs have been used in several applications [3, 6, 33, 36-40], and they often outperform their type-1 (T1) counterparts when handling uncertainties. The output of the inference engine of an IT2 FLS is an IT2 FS. It requires conversion to a T1 FS before the final crisp output can be calculated. This process is called type-reduction (TR) [7, 8]. Originally, TR was performed using the iterative Karnik-Mendel (KM) algorithms [9], which are computationally intensive. Several methods were introduced to enhance the computational performance of KM algorithms (see for example, [10-13, 35] and references therein). Other methods proposed non-iterative alternatives to TR. A summary of several alternative approaches is provided in [14], and a comparison of their computational cost in [13]. Both categories of methods have demonstrated good performance; and the question of which category is better remains an open problem [14].

Having a closed-form fuzzy inference engine relationship is preferred in control design, especially when stability analysis is required [41-43]. The iterative KM algorithms cannot be suitable in providing such a form; therefore, alternative approaches must be utilized [44]. To the best of the authors' knowledge, none of the alternative methods directly provides a simple closed mathematical form that fully represents an IT2 FLS. In [44], the authors developed an inference mechanism for an IT2 Takagi–Sugeno–Kang fuzzy logic control system that has a closed-mathematical form when antecedents are type-2 fuzzy sets and consequents are crisp numbers. It was based on the Wu-Mendel Uncertainty Bounds method.

In this paper, we develop two alternative closed-form representations for IT2 FLSs. The two representations are based on approximating two existing methods, namely, Coupland and John's Geometric Centroid, and the Nie-Tan method. Coupland and John's Geometric Centroid (GC) defuzzification [15-21] provides a good approximation to the type-reduced Centroid [22]. TR is by-passed by directly finding the x-coordinate of the geometric centroid of the FOU of the output of the IT2 FS. One of the characteristics of using the GC method is that the two implication operations performed to find the upper and lower bounds of the resulting IT2 FS are completely independent from one another and can be performed concurrently [15], and type-1 fuzzy operations can be utilized in doing so.

We introduce an approach to reach a closed-form mathematical representation for IT2 FLSs based on the GC method. We also extend the results to provide another closed-form mathematical representation based on the Nie-Tan operator [30]. The rest of the paper is organized as follows. In section 2 we provide background information on IT2 FLSs with detailed review of the GC method along with the mathematical formulations. Section 3 introduces the closed form representations. Section 4 provides simulation results of the outputs of the two introduced systems, followed by the conclusions in section 5. Summary of the symbols and the abbreviations used in the paper, and their meanings is provided in the appendix.

## 2. Background

Consider a two-input single-output IT2 FLS. The domain of each input is partitioned into $N$ IT2 FSs. There are $M = N^2$ possible rules in the rule-base. The $k^{th}$ rule (denoted $R^k$) is given by;

$R^k$: IF $x_1$ is $\tilde{F}_1^i$ AND $x_2$ is $\tilde{F}_2^j$ THEN $y$ is $\tilde{G}^{ij}$; $(i, j = 1, 2, ..., N)$     (1)

where $k = (i-1)N + j$, $k \in [1, M]$, [3]. The $k^{th}$ rule can also be referred to as the $ij^{th}$ rule. The rules consequents $\tilde{G}^{ij}$ are, in general, IT2 FSs, but they can be intervals or type-1 fuzzy sets. For singleton fuzzification, the firing set (firing level) of $R^k$ is given by (2).

$$F^{ij}(x) = \mathcal{T}\left(\mu_{\tilde{F}_1^i}(x_1), \mu_{\tilde{F}_2^j}(x_2)\right) = \mu_{\tilde{F}_1^i}(x_1) \star \mu_{\tilde{F}_2^j}(x_2) \quad (2)$$

where both $\mathcal{T}(.)$ and '$\star$' indicate the utilized t-norm operator (commonly, the *minimum* or *product* operator). It follows that:

$$F^{ij}(x) = \left[\underline{\mu}_{\tilde{F}_1^i}(x_1) \star \underline{\mu}_{\tilde{F}_2^j}(x_2), \overline{\mu}_{\tilde{F}_1^i}(x_1) \star \overline{\mu}_{\tilde{F}_2^j}(x_2)\right] \equiv [\underline{f}^{ij}, \overline{f}^{ij}] \quad (3)$$

where $\underline{f}^{ij}$ and $\overline{f}^{ij}$ are the lower and upper firing degrees of the $k^{th}$ rule, and $\underline{\mu}_{\tilde{F}^i}(x)$ and $\overline{\mu}_{\tilde{F}^i}(x)$ are the lower and upper membership grades of $\tilde{F}^i(x)$. When the consequent of $R^k$ is a T1 FS, $G^k$ [15], then, the implied fuzzy set is given by:

$$\mu_{\tilde{B}^k}(y) = \left[\underline{\mu}_{\tilde{B}}(y), \overline{\mu}_{\tilde{B}}(y)\right] = [\underline{f}^{ij} \star \mu_{G^{ij}}(y), \overline{f}^{ij} \star \mu_{G^{ij}}(y)] \quad (4)$$

Both the upper and lower bounds, in the previous steps, are reached independently using type-1 operations. The next step is to find the overall output IT2 FS by aggregating the implied sets from all fired rules. Assuming that all rules are being fired (the non-fired rules are considered to be fired with zero firing strengths), the output fuzzy set is given by;

$$\mu_{\tilde{B}}(y) = \sqcup_{k=1}^{M} \mu_{\tilde{B}^k}(y) = \left[\sqcup_{k=1}^{M}(\underline{f}^k \star \mu_{G^k}(y)), \sqcup_{k=1}^{M}(\overline{f}^k \star \mu_{G^k}(y))\right] \quad (5)$$

Here we use "$\sqcup$" to represent the aggregation operation. Since we use type-1 operations to find each of the two bounds, it is natural to utilize type-1 aggregation operators (*max*, *sum*, or *probabilistic or* are the common ones used with T1 FSs [23]).

In the GC method, type-reduction is bypassed by finding the center of area (CoA) of the FOU of the output T2 FS (the area bounded by the upper and lower output membership functions (UMF and LMF), $\underline{\mu}_{\tilde{B}}(y)$ and $\overline{\mu}_{\tilde{B}}(y)$), an approach assumed to be natural extension to commonly using the center of area defuzzifier in T1 FS [20]. The upper and lower membership functions (UMF($\tilde{B}$) and LMF($\tilde{B}$)) are respectively given by (6) and (7).

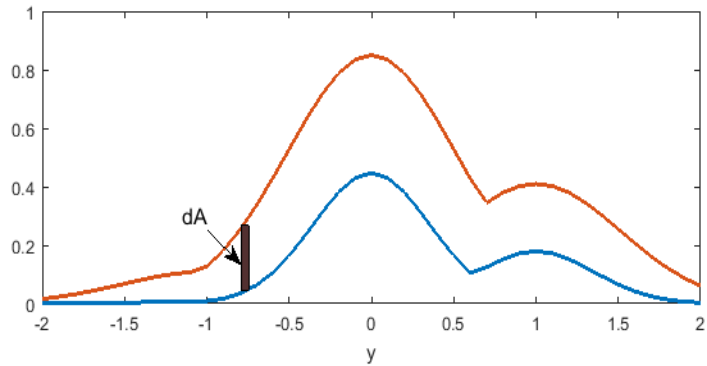

Fig. 1. FOU of a sample output IT2 FS.

$$\bar{\mu}_{\tilde{B}}(y) = \sqcup_{k=1}^{M}(\bar{f}^k \star \mu_{G^k}(y)) = \sqcup_{k=1}^{M}(\bar{\mu}_{\tilde{F}_1^i}(x_1) \star \bar{\mu}_{\tilde{F}_2^j}(x_2) \star \mu_{G^k}(y)) \quad (6)$$

$$\underline{\mu}_{\tilde{B}}(y) = \sqcup_{k=1}^{M}(\underline{f}^k \star \mu_{G^k}(y)) = \sqcup_{k=1}^{M}(\underline{\mu}_{\tilde{F}_1^i}(x_1) \star \underline{\mu}_{\tilde{F}_2^j}(x_2) \star \mu_{G^k}(y)) \quad (7)$$

Mathematically, the area of the region bounded by the *UMF* and *LMF* is found by integrating *dA* with respect to the output variable *y* (see Fig. 1). The centre of area can be found [24] using the expression in Eq. (9).

$$dA = \left(\bar{\mu}_{\tilde{B}}(y) - \underline{\mu}_{\tilde{B}}(y)\right) \cdot dy \quad (8)$$

$$y_{Crisp} = CoA(\tilde{B}) = \frac{\int_{-\infty}^{\infty} y \cdot dA}{\int_{-\infty}^{\infty} dA} = \frac{\int_{-\infty}^{\infty} y \cdot (\bar{\mu}_{\tilde{B}}(y) - \underline{\mu}_{\tilde{B}}(y)) \cdot dy}{\int_{-\infty}^{\infty} (\bar{\mu}_{\tilde{B}}(y) - \underline{\mu}_{\tilde{B}}(y)) \cdot dy} \quad (9)$$

When the output membership functions of the fuzzy systems are T1 sets, Eq. (9) can be simplified to;

$$\frac{\int_{-\infty}^{\infty} y \cdot (\bar{\mu}_{\tilde{B}}(y)) \cdot dy - \int_{-\infty}^{\infty} y \cdot (\underline{\mu}_{\tilde{B}}(y)) \cdot dy}{\int_{-\infty}^{\infty} (\bar{\mu}_{\tilde{B}}(y) - \underline{\mu}_{\tilde{B}}(y)) \cdot dy} = \frac{C_u A_u - C_l A_l}{A_u - A_l} \quad (10)$$

where $C_u$, $C_l$, $A_u$ and $A_l$ respectively represent the centers and the areas of the upper and lower membership functions of $\tilde{B}(y)$ [26-28].

The integration process involved in (9) is computationally intensive. An approach to reduce the complexity, as in T1 FLSs [15], is to replace the integration process by using discretized weighted average operation (discretized in *n* points). In this case, Eq. (9) can be rewritten as:

$$CoA(\tilde{B}) \cong \frac{\sum_{i=1}^{n} y_i \mu_{\tilde{B}}(y_i)}{\sum_{i=1}^{n} \mu_{\tilde{B}}(y_i)} \quad (11)$$

Alternatively, in John and Coupland's Geometric Centroid method [15, 21], the integration is replaced by using simple geometry to replicate the weighted average operation. The FOU is first approximated by regular polygons, mostly triangles, and then the weighted average of the polygons is calculated. Operations such as join and meet are carried out using methods from computational geometry. In the following section we introduce approximations to the above discussed expressions to reach the required representation of the IT2 FS.

## 3. Closed Form Representation

Equation (11) provides an expression that represents the output of an IT2 FLS, using the CoA of the output FOU as an IT2 defuzzification method. As discussed in the background section, the arithmetic steps involved in computing both the upper and lower MFs were mostly independent of each other. Type-1 operations are utilized in reaching both the UMF and LMF. Using the same reasoning, we can use other T1 techniques to further simplify the representation of the IT2 FLS.

## 3.A. Derivation of the Expressions

When using center average defuzzification the shape of the membership functions associated with the output fuzzy sets does not affect the final output. Therefore, one can use singletons centered at the appropriate positions [29]. In (8) and (9), if sum operator is used for join, product for t-norm, and utilizing singleton fuzzy membership functions $\mu_{G^k}(y) = \delta(y - b_k)$ for the outputs ($G^k$), where the '$\delta$' is the Dirac delta function, used to represent the fuzzy singleton, and $b_k$ are the locations of the output membership functions (this is equivalent to using center-average defuzzification, with $b_k$ representing the centers of the output membership functions). Substituting in (11), the output is rewritten as:

$$y_{crisp} = \frac{\sum_{k=1}^{M} b_k \left( \bar{\mu}_{\tilde{F}_1^i}(x_1) \cdot \bar{\mu}_{\tilde{F}_2^j}(x_2) \right) - \sum_{k=1}^{M} b_k \left( \underline{\mu}_{\tilde{F}_1^i}(x_1) \cdot \underline{\mu}_{\tilde{F}_2^j}(x_2) \right)}{\sum_{k=1}^{M} \bar{\mu}_{\tilde{F}_1^i}(x_1) \cdot \bar{\mu}_{\tilde{F}_2^j}(x_2) - \sum_{k=1}^{M} \underline{\mu}_{\tilde{F}_1^i}(x_1) \cdot \underline{\mu}_{\tilde{F}_2^j}(x_2)} \tag{12}$$

Equation (12) can be represented in a more compact form:

$$y_{crisp} = \frac{\sum_{k=1}^{M} b_k \left( \bar{\mu}_{\tilde{F}_1^i}(x_1) \cdot \bar{\mu}_{\tilde{F}_2^j}(x_2) - \underline{\mu}_{\tilde{F}_1^i}(x_1) \cdot \underline{\mu}_{\tilde{F}_2^j}(x_2) \right)}{\sum_{k=1}^{M} (\bar{\mu}_{\tilde{F}_1^i}(x_1) \cdot \bar{\mu}_{\tilde{F}_2^j}(x_2) - \underline{\mu}_{\tilde{F}_1^i}(x_1) \cdot \underline{\mu}_{\tilde{F}_2^j}(x_2))} \tag{13}$$

It is understood that carefulness is required when designing the fuzzy system, so that the denominator of (13) is not equal to zero for any value of the inputs. Also, if different output singletons are used for upper and lower, the crisp output of (12) can be re-written as:

$$y_{crisp} = \frac{\sum_{k=1}^{M} \bar{b}_k \left( \bar{\mu}_{\tilde{F}_1^i}(x_1) \cdot \bar{\mu}_{\tilde{F}_2^j}(x_2) \right) - \sum_{k=1}^{M} \underline{b}_k \left( \underline{\mu}_{\tilde{F}_1^i}(x_1) \cdot \underline{\mu}_{\tilde{F}_2^j}(x_2) \right)}{\sum_{k=1}^{M} (\bar{\mu}_{\tilde{F}_1^i}(x_1) \cdot \bar{\mu}_{\tilde{F}_2^j}(x_2) - \underline{\mu}_{\tilde{F}_1^i}(x_1) \cdot \underline{\mu}_{\tilde{F}_2^j}(x_2))} \tag{14}$$

Equation (13) represents a whole IT2 FLS in a closed-form, similar to its T1 counterpart [29].

An alternative closed-form mathematical expression can be obtained by utilizing the Nie-Tan operator [30]. It was recently shown [31] that the closed-form Nie-Tan (NT) operator is an accurate IT2 defuzzifing method. The method outputs a T1 FS that is the average of the upper and lower bounds of the FOU of the output IT2 FS '$\tilde{A}$', as given by Eq. (15). The final system output is then found by calculating the center-of-gravity of $\tilde{A}$, which can be computed by (16).

$$\mu^*(y) = \frac{1}{2}(\bar{\mu}(y) + \underline{\mu}(y)) \tag{15}$$

$$C(\tilde{A}) = \frac{\int_{y_{min}}^{y_{max}} \mu^*(y) \cdot y \, dy}{\int_{y_{min}}^{y_{max}} \mu^*(y) \, dy} \tag{16}$$

Using (15) and following similar approach to our earlier discussion, we can reach another closed-form approximation for the IT2 FLS, as given in (17), based on the Nie-Tan method.

$$y_{crisp} = \frac{\sum_{k=1}^{M} b_k \left( \bar{\mu}_{\tilde{F}_1^i}(x_1) \cdot \bar{\mu}_{\tilde{F}_2^j}(x_2) + \underline{\mu}_{\tilde{F}_1^i}(x_1) \cdot \underline{\mu}_{\tilde{F}_2^j}(x_2) \right)}{\sum_{k=1}^{M} (\bar{\mu}_{\tilde{F}_1^i}(x_1) \cdot \bar{\mu}_{\tilde{F}_2^j}(x_2) + \underline{\mu}_{\tilde{F}_1^i}(x_1) \cdot \underline{\mu}_{\tilde{F}_2^j}(x_2))} \tag{17}$$

*3.B. Selection of Membership Functions*

In principle, any form of membership function can be used with the developed fuzzy systems, represented by equations (13) and (17). To have a true mathematical representation (i.e. to avoid the discontinuities associated with expressing membership functions such as trapezoidal, triangular, etc.), we limit the membership functions to be Gaussian;

$$\mu_{F^i}(x) = \exp\left(-\frac{1}{2}\left(\frac{x-m_i}{\sigma_i}\right)^2\right) \quad (18)$$

When the utilized IT2 membership functions have certain means and uncertain variances, both UMFs and LMFs are simpler to represent [32]. On the other hand, when they have uncertain means, it becomes necessary to redefine the UMFs and LMFs to avoid using *max* and *min* operators. As shown in fig. 2, to circumvent this we may approximate both the UMF and LMF with Gaussian equivalents using curve fitting techniques. In the following section we examine both presented IT2 systems. In this figure, the LMF is exactly described by (19), while the approximate LMF is described by (20). Similar expressions can be written for the UMF.

$$\underline{\mu}(x) = \min\left(\exp\left(-\frac{1}{2}\left(\frac{x+0.1}{0.418}\right)^2\right), \exp\left(-\frac{1}{2}\left(\frac{x-0.1}{0.418}\right)^2\right)\right) \quad (19)$$

$$\underline{\mu}(x) \approx 0.9183 \exp\left(-\frac{1}{2}\left(\frac{x}{0.3651}\right)^2\right) \quad (20)$$

To summarize this section, equations (13) and (17) are introduced to provide two different closed-form simple representations of IT2 FLSs. The main difference between the two equations is that each of them is based on a different IT2 FLS architecture. Eq. (13) is based on "Geometric interval type-2 systems" approach, initially introduced by Coupland and John in Ref. [18]. Their method approximates the crisp output of the system by computing the geometric center (GC) of the footprint of uncertainty (FOU) of the output IT2 FS. Eq. (13) is a simplified representation (single closed-form expression) of their method.

On the other hand, Eq. (17) is based on the Nie-Tan approach, originally introduced by Nie and Tan in [30] and further investigated by John, Coupland and Kendall in [31]. Contrary to the GC method, the Nie-Tan method computes a representative type-1 fuzzy set (T1 FS) that is an average of the upper and lower membership functions of the output IT2 FS. The crisp output is computed as the center-of-gravity of this representative T1 set. Eq. (17) is a simplified representation of the NT method. We also used approximate Gaussian forms for both the UMF and LMF to simplify the representation when the utilized IT2 membership functions have uncertain means.

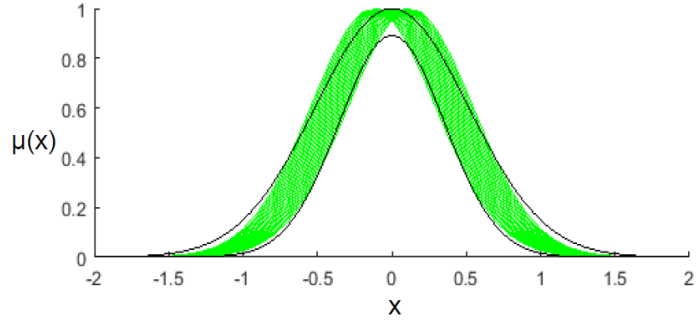

Fig. 2. An example of a Gaussian membership function with uncertain mean ∈ $[\mu - \Delta\mu, \mu + \Delta\mu]$, and fixed standard deviation σ (green shaded area, here μ = 0, $\Delta\mu = 0.1$, and σ = 0.418), and the approximation of its UMF and LMF (black solid lines; the approximate UMF is Gaussian with μ = 0, and σ = 0.4937, the LMF is a scaled-down Gaussian, with μ = 0, σ = 0.3651, and a scaling factor of 0.9183).

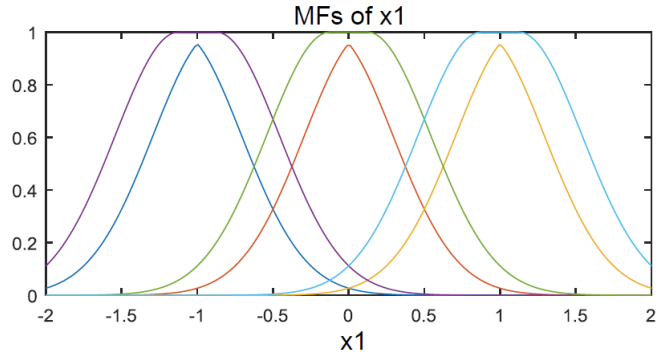

Fig. 3. The three membership functions for one of the inputs $x_1$. By limiting the range of the input to values ∈[-1, 1], the left and right membership functions are practically S and Z Gaussian. For the three membership functions, (the means are $\mu_1 = -1$, $\mu_2 = 0$, $\mu_3 = 1$, $\Delta\mu = 1/8$, and σ = 0.418), and the approximation of its UMF and LMF (black solid lines; the approximate UMF is Gaussian with μ = 0, and σ = 0.5128, the LMF is a scaled-down Gaussian, with μ = 0, 0.3532, and a scaling factor of 0.895.

## 4. Results and discussions

In this section we first show how the proposed equations can be used to design an IT2 FLS, and then we show how they perform in stabilizing an inverted pendulum.

### 4.A. Control Surface

We used (13) and (17) to represent a simple two-input single-output IT2 fuzzy system. The system utilizes three membership functions in the domain of each input. Fig. 3 illustrates the three membership functions utilized in one of the input domains (both are identical). The rule-base used is shown in table 1, with the corresponding values of $b_i$'s. The three membership functions have uncertain means that vary between $\mu_i - \Delta\mu$ and $\mu_i + \Delta\mu$ ($\mu_1 = -1$, $\mu_2 = 0$, $\mu_3 = 1$, $\Delta\mu = 1/8$, and σ = 0.418). To use (13) and (17), each UMF and LMF was approximated using Gaussian functions, as described in the previous section. The approximate UMFs are as follows; (with the same values of $\mu_i$) with σ = 0.5128. Each LMF is a scaled-down Gaussian with σ = 0.3532, and an amplitude scaling factor of 0.895.

Figures 4 to 7 show the resulting surfaces of the introduced systems of (13) and (17) with both exact and approximate forms of UMFs and LMFs.

TABLE 1. RULE-BASE FOR THREE MEMBERSHIP FUNCTIONS

| Rule number | 1st input | 2nd input | Output | Corresponding bi value |
|---|---|---|---|---|
| 1 | N | N | P | 1 |
| 2 | N | Z | P | 1 |
| 3 | N | P | Z | 0 |
| 4 | Z | N | P | 1 |
| 5 | Z | Z | Z | 0 |
| 6 | Z | P | N | -1 |
| 7 | P | N | Z | 0 |
| 8 | P | Z | N | -1 |
| 9 | P | P | N | -1 |

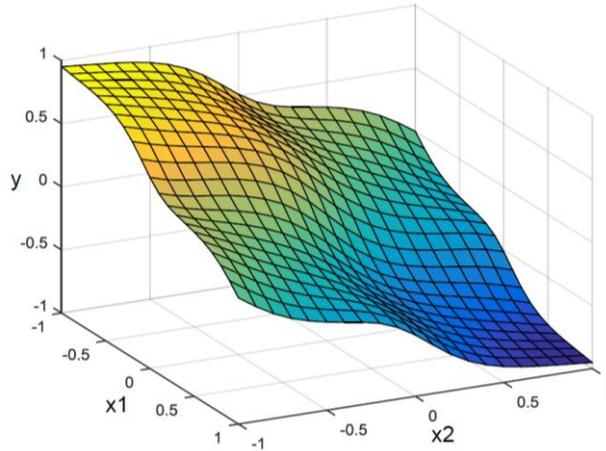

Fig. 4. Surface using the introduced closed-form mathematical representation for (13), with the approximated Gaussian functions.

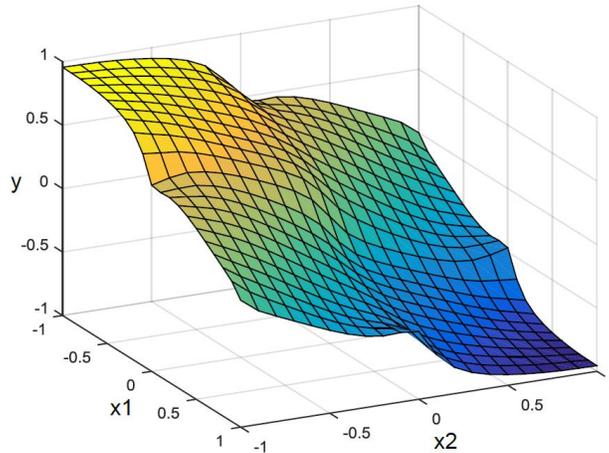

Fig. 5. Surface using the introduced closed-form mathematical representation for (13), with the exact representations of UMF and LMF.

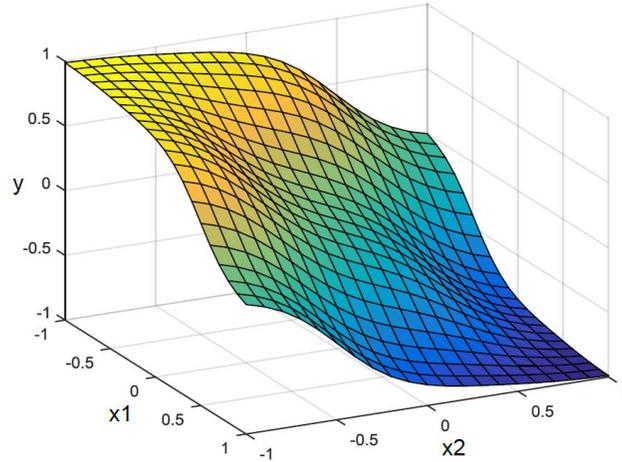

Fig.6. Surface using the closed-form mathematical Nie-Tan representation for (17), with the approximated Gaussian functions.

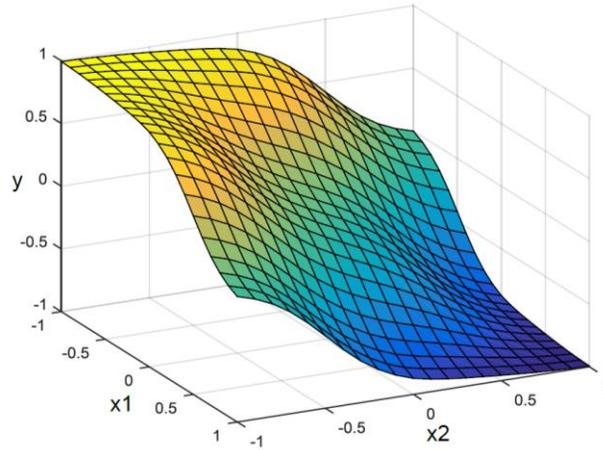

Fig. 7. Surface using the closed-form mathematical Nie-Tan representation for (17), with the exact representations of UMF and LMF.

### *4.B. Application Example: Control of an Inverted Pendulum*

Balancing an inverted pendulum is a common control problem. We utilize the two introduced equations (i.e. (13) and (17)) in a feedback control scheme to balance an inverted pendulum and compare the performance to that of the two reference IT2 methods (i.e. NT and GC).

An inverted pendulum mounted on a cart is balanced by applying dynamic force $f$ to the cart. It is common to use the error signal (the angular difference between the upright position and the actual position of the pendulum) and its derivative as inputs to the feedback controller utilized in balancing the pendulum. Figure 8 shows a block diagram of the control system. As discussed earlier, the gains are introduced to scale the universes of discourse for the fuzzy membership functions. Here, the values for $g_1$, $g_2$, and $g_y$ are $4/\pi$, $0.4/\pi$, and 100, respectively. We use the model described by (21) and (22) for the inverted pendulum system [29].

$$\ddot{y} = \frac{\left(g\sin(y) + \cos(y)\left[\dfrac{-\bar{f} - 0.25\dot{y}^2\sin(y)}{1.5}\right]\right)}{\left(\dfrac{2}{3} - \dfrac{1}{6}\cos^2(y)\right)} \tag{21}$$

$$\dot{\bar{f}} = -100\bar{f} + 100f \tag{22}$$

where, $f$ is the applied force in Newton, $y$ is the angular position in radians, and $g$ is the gravity of Earth.

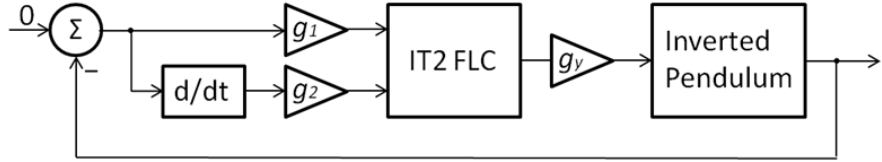

Fig. 8. Block-diagram model of inverted pendulum control.

Figure 9 shows the MATLAB simulation results of controlling this inverted pendulum utilizing the GC method (numerically evaluated), and utilizing the two different realizations of (13) discussed earlier (i.e. using exact Gaussian membership functions, and using the approximate form). Figure 10 shows similar results when the NT method and (17) are utilized. Both figures show that the introduced formulas performed very closely to their more complexly-represented counterparts.

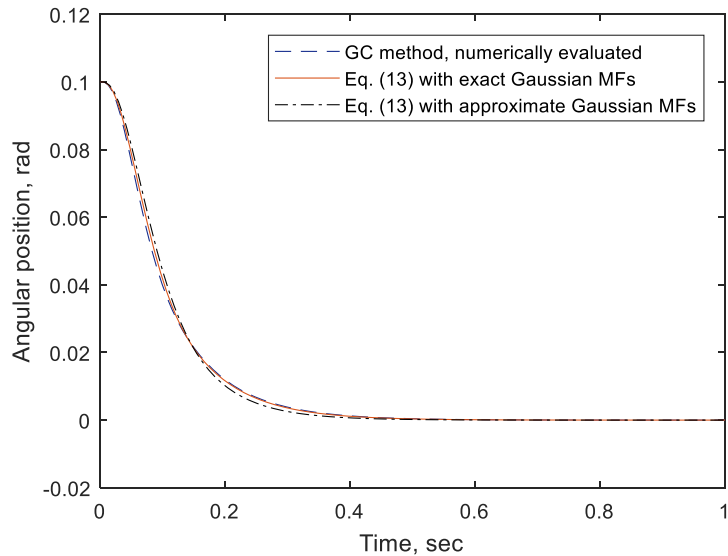

Fig. 9. Error in angular position of an inverted pendulum controlled by the GC method, and by the proposed eq. (13)

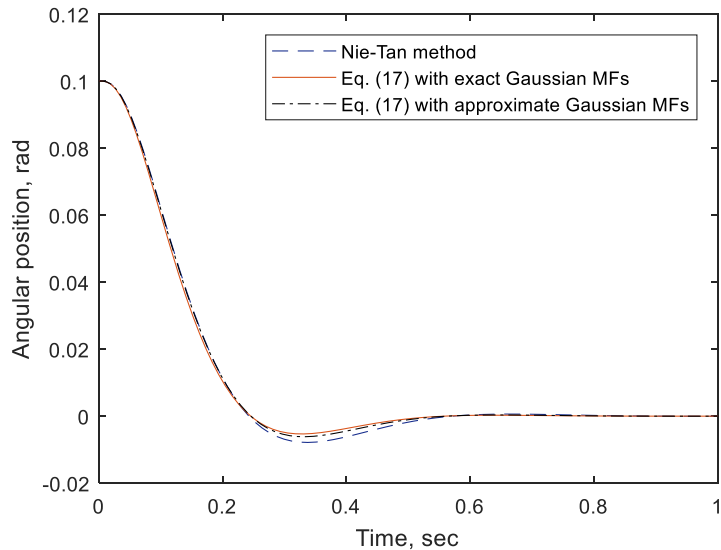

Fig. 10. Error in angular position of an inverted pendulum controlled by the NT method, and by the proposed eq. (17).

## 5. Conclusions

The paper presented an approach for reaching closed-form mathematical representations of interval type-2 fuzzy logic systems based on the notation that both the UMF and LMF can be found independently. First, by using the CoA of the FOU as an approximation of the crisp output of an IT2 FLS, a mathematical expression is reached. The same approach was utilized with the Nie-Tan operator, resulting in a different closed-form mathematical expression. Simulation results are introduced to show the resulting IT2 fuzzy output surfaces. Both expressions provide simpler methods of utilizing and studying IT2 FLSs, their performance in control system is an open research problem, as is the case for the several previously presented alternative representations of T2 FLSs.

# Appendix A.

Table A1. Abbreviations and their meanings

| Abbreviation | Meaning |
| --- | --- |
| CoA | Center of area |
| FOU | Footprint of uncertainty |
| GC | Geometric Centroid |
| IT2 FLS | Interval type-2 fuzzy logic system |
| IT2 FS | Interval type-2 fuzzy set |
| IT2 TSK FLC | Interval type-2 Takagi–Sugeno–Kang fuzzy logic controller |
| KM | Karnik-Mendel algorithms |
| LMF | Lower membership function |
| MF | Membership function |
| NT | Nie-Tan operator |
| T1 | Type-1 |
| T2 | Type-2 |
| T1 FS | Type-1 fuzzy set |
| T2 FS | Type-2 fuzzy set |
| TR | Type-reduction |
| UMF | Upper membership function |

Table A2. Symbols and their meanings

| Symbol | Meaning |
| --- | --- |
| $\tilde{A}$ | Type-2 fuzzy set |
| $A_l$ | The area under the LMF of the output IT2 FS $\tilde{B}(y)$ |
| $A_u$ | The area under the UMF of the output IT2 FS $\tilde{B}(y)$ |
| $b_k$ | The locations of the output membership functions |
| $C_l$ | The geometric center of the LMF of the output IT2 FS $\tilde{B}(y)$ |
| $C_u$ | The geometric center of the UMF of the output IT2 FS $\tilde{B}(y)$ |
| $\delta$ | The Dirac delta function (used to represent the fuzzy singleton) |
| $F^{ij}(x)$ | The firing set (firing level) of the $k^{th}$ (the $ij^{th}$) rule |
| $\underline{f}^{ij}$ | The lower firing degree of the $k^{th}$ (the $ij^{th}$) rule |
| $\overline{f}^{ij}$ | The upper firing degree of the $k^{th}$ (the $ij^{th}$) rule |
| $R^k$ or $R^{ij}$ | The $k^{th}$ rule of the rule-base |
| $\mathcal{T}(.)$ and '★' | The T-norm operator (commonly the minimum or product) |
| $\bar{\mu}_{\tilde{F}^i}(x)$ | The upper membership grades of $\tilde{F}^i(x)$. |
| $\underline{\mu}_{\tilde{F}^i}(x)$ | The lower membership grade of $\tilde{F}^i(x)$. |
| $\mu_{\tilde{B}}(y)$ | The membership grade of the IT2 FS $\tilde{B}$ |
| $\underline{\mu}_{\tilde{B}}(y)$ | The lower membership function LMF($\tilde{B}$) |
| $\bar{\mu}_{\tilde{B}}(y)$ | The upper membership function UMF($\tilde{B}$) |
| $\sqcup$ | The join operation |